\documentclass{sig-alternate}

\usepackage[
  pass,
]{geometry}

\usepackage{capt-of}

\usepackage{xspace}
\usepackage{booktabs} 
\usepackage{url}

\usepackage{color}
\usepackage{enumitem} 
\newcommand{\system}{Carina\xspace}
\usepackage{hyperref}
\hypersetup{
colorlinks = true,
linkcolor = blue,
urlcolor  = blue,
citecolor = blue,
anchorcolor = blue
}

\newcommand{\hide}[1]{}

\permission{\copyright 2017 International World Wide Web Conference Committee (IW3C2), 
published under Creative Commons CC BY 4.0 License.}
\conferenceinfo{WWW 2017,}{April 3--7, 2017, Perth, Australia.}
\copyrightetc{ACM \the\acmcopyr}
\crdata{978-1-4503-4913-0/17/04. \\
http://dx.doi.org/10.1145/3041021.3054234 \\
\includegraphics{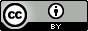}
}
 
\begin{document}

\title{\system: Interactive Million-Node Graph Visualization using 
Web Browser
Technologies}
%
%
%
%
%

\numberofauthors{1} 
%
\author{
%
%
\alignauthor
Dezhi Fang, Matthew Keezer, Jacob Williams, Kshitij Kulkarni,\\ Robert Pienta, Duen Horng Chau\\
       \affaddr{Georgia Institute of Technology}\\
       \email{\{dezhifang, rkeezer3, jake.williams, kkulkarni36, pientars, polo\}@gatech.edu}
}

\makeatletter
\let\@oldmaketitle\@maketitle
\renewcommand{\@maketitle}{\@oldmaketitle
	\vspace{-0.25in}
	\centering
	\includegraphics[width=0.90\textwidth]{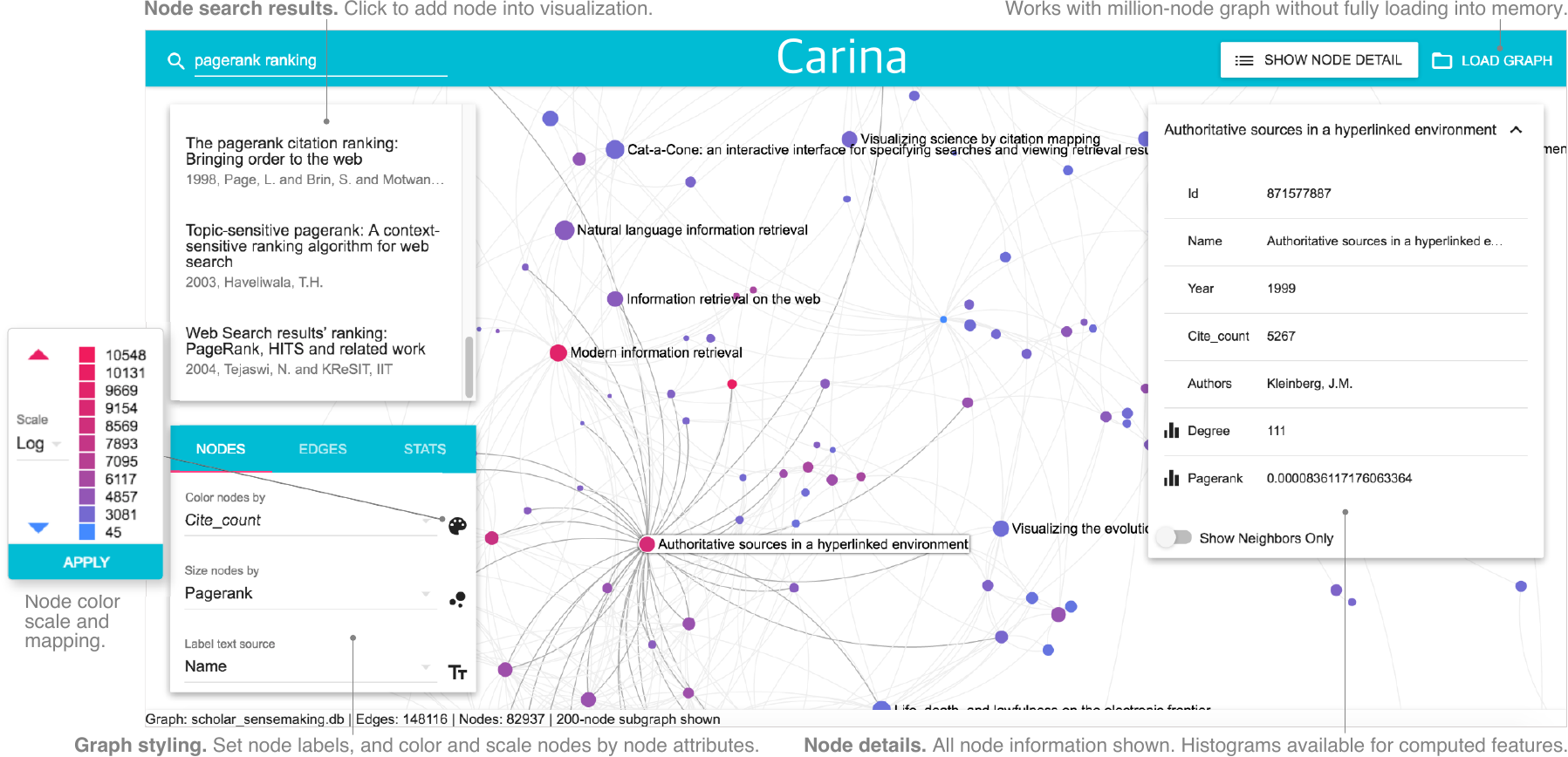}
\vspace{-0.1in}
\captionof{figure}{\system visualizing citation network data (83k nodes, 150k edges) \cite{chau2011apolo}. 
	\system works with out-of-core million-node graphs (up to 69M edges).
	\system uses latest web browser technologies for high-performance graph rendering (WebGL), easy cross-platform deployment (Electron), and lightweight, scalable data storage.}
\vspace{0.05in}
}
\label{fig:crownjewel}
\makeatother

\maketitle


\begin{abstract}
We are working on a scalable, interactive visualization system, called \system, for people to explore million-node graphs.
By using latest web browser technologies, \system offers fast graph rendering via WebGL and works across desktop (via Electron) and mobile platforms. 
Different from most existing graph visualization tools, \system does not store the full graph in RAM,
enabling it to work with graphs with up to 69M edges.
We are working to improve and open-source \system, to offer researchers and practitioners a new, scalable way to explore and visualize large graph datasets.
\end{abstract}



\section{Introduction}
Large graph data have become increasingly common.
Visualizing graphs can help people more easily understand relations among entities.
However, existing graph visualization systems face  multiple technical, visual, and scalability challenges.

\vspace{0.05in}
\noindent \textit{Visual Scalability Challenge.}
Most graph visualization systems follow the convention approach of visualizing the entire graph (e.g., Gephi \cite{bastian2009gephi}, Cytoscape \cite{shannon2003cytoscape}). For million-node graphs, such visualizations generate  ``hairballs'' with extreme edge crossings \cite{chau2011apolo}, overwhelming human perception and impeding understanding. 

\vspace{0.05in}
\noindent \textit{Data Scalability Challenge.}
In most existing systems, a graph dataset must first be fully loaded into memory. 
The available RAM becomes a barrier for analyzing larger graphs with million nodes or more.
For example, the popular Gephi system runs out of RAM when trying to load the LiveJournal social graph with 69M edges \cite{leskovec2015snap}.

\vspace{0.05in}
\noindent \textit{Technology Challenge.}
Most existing visualization tools are built primarily for desktop use,
precluding analysis in mobile environments, which are increasingly common (e.g., DARPA GUARD DOG mobile graph analytics \cite{guarddog2010darpa}).
For instance, 
Gephi and Cytoscape are built using Java and run only on desktop computers.



We present our ongoing research to tackle the above challenges.

\begin{itemize}[leftmargin=0.7cm, labelindent=-1cm, itemsep=0cm, topsep=-0.15cm]

\item 
We introduce \system (Fig. \ref{fig:crownjewel}), a cross-platform prototype graph visualization system.
Using latest web browser technologies,
\system offers fast graph rendering via WebGL, and lightweight, cross-platform deployment via Electron.


\item 
We demonstrate that \system works with million-node graphs (up to 69M edges), without requiring them to fit in RAM, unlike most existing graph visualization tools.
\system visualizes user-specified subgraphs (instead of the whole graph), allowing users to focus their exploration on the most relevant graph regions.
\system uses SQLite for storing and querying out-of-core datasets.

\end{itemize}

\section{CARINA Feature Highlights}

\hide{ARGO's architecture is split into three parts: Backend, Graph Renderer, and User Interaction.}

\noindent \textbf{High-performance Graph Rendering via WebGL.}
Many visualization libraries, e.g., d3.js ({\small\url{https://d3js.org}}) and processing.js, 
render graphs using technologies like \textit{Scalable Vector Graphics} (SVG) and HTML Canvas elements.
However, their rendering speed 
(screen refresh rate) begins to deteriorate significantly starting at low hundreds of nodes and edges
({\small\url{https://github.com/anvaka/graph-drawing-libraries}}).
%

We have identified WebGL as a high-performance graphics technology that has strong potential for fast graph rendering.
\mbox{WebGL} uses GPU acceleration,
and is supported by all modern web browsers ({\small\url{http://caniuse.com/#feat=webgl}}).
%
\system leverages WebGL's high-speed graphics capabilities through Three.js, a higher-level library designed to simplify WebGL programming ({\small\url{https://threejs.org}}). 
\system adapts many 3D accelerated graphics techniques from WebGL to 2D space for graph visualization. 
In particular, we use level-of-detail, buffer geometry, and particle systems for fast rendering of nodes and edges.

\textit{Visual Scalability on Real Data.}
To better understand WebGL's rendering scalability,
we tested \system with graphs of varying sizes, on a MacBook Pro laptop (2015 version, i7-4870HQ, 16GB RAM).
For the YEAST dataset with 2361 nodes and 7182 edges, user interactions such as panning, zooming and dragging nodes, with a force-directed graph layout algorithm running in the background, 
achieved a smooth frame rate of 60FPS.
The frame rate only starts to drop below 30FPS when the graph size exceeds 20k nodes and 56k edges.
We note that when analyzing real-world power-law graphs, we typically would not want to visualize the entire graph (which shows up as a ``hairball'').
We conducted the above experiment to understand the technological limits.


\vspace{0.05in}
\noindent \textbf{Cross-platform Integration and Deployment.}
%
We design \system with platform portability in mind, hence our decision to use latest web browser technologies.
\system can be deployed as a desktop application that runs on all popular operating systems (Linux, Windows, Mac), via the Electron framework based on the Chrome browser and Node.js ({\small\url{http://electron.atom.io/}}).
Electron packages \system as  self-contained binaries for all platforms without assumptions of the user's run-time environment.
Electron also grants \system native I/O and high performance Inter-Process Communication (IPC) capabilities.  
\system  can also be used in any modern browsers, allowing it to run on  mobile devices and  desktops. 


\vspace{0.05in}
\noindent \textbf{Million-node Graph Data Storage and Exploration.}
Different from most graph visualization tools, 
\system does not store the full graph in RAM; only the visualized subgraph is kept in memory,
allowing \system to work with graphs with up to 69M edges \cite{leskovec2015snap}.
Currently, \system stores a million-node graph in an out-of-core SQLite database.
We chose SQLite for its simplicity, ease of integration, and scalability for up to tens of millions of edges.
The user can select and visualize sub-graphs through techniques such as node search (as in Fig. \ref{fig:crownjewel}) or based on graph features (e.g., computed measures like PageRank scores).
SQLite can induce subgraphs quickly. For example, it takes only 120ms to induce a 2000-node subgraph (9867 edges) out of the 69M edge LiveJournal graph.


\section{Conclusions \& Ongoing Work}
We are designing and developing \system, a scalable, interactive visualization system for million-node  graph exploration.
Using latest web browser technologies,
\system offers multiple advantages over existing graph visualization tools, such as fast graph rendering (via WebGL), easy cross-platform deployment (via Electron), and scalable data storage and exploration of large graphs with up to 69M edges on commodity machines.
We plan to conduct lab studies to evaluate the usability of \system, and work with real domain users, such as security analysts at Symantec, to use \system to help uncover company insider threats lurking in large email graphs and computer communication networks.
We believe \system provides a new, scalable way for practitioners and researchers to explore and visualize large graph datasets.

\section*{Acknowledgement}
We thank Andrei Kashcha for the WebGL inspiration for graphs.
This work is supported by NSF IIS-1563816, 1217559, TWC-1526254.

\bibliographystyle{abbrv}
\bibliography{sigproc} 
\end{document}